\documentclass[aps,prl,nofootinbib,showpacs,twocolumn]{revtex4-1}
\usepackage{graphicx}
\usepackage{dcolumn}
\usepackage{amsmath}
\usepackage{latexsym}
\usepackage{mathrsfs}
\usepackage{color}
\def\clrr {\color{black}}

\begin{document}

\title{Universality  splitting  in distribution of number of miRNA  co-targets}

\author{Mahashweta Basu$^1$, Nitai P. Bhattacharyya$^2$, P. K. Mohanty$^1$}
\affiliation{ $^1$Condensed Matter Physics Division, Saha Institute of Nuclear Physics, 1/AF Bidhan Nagar, Kolkata 700064, India.\\
$^2$Crystallography and Molecular Biology Division, Saha Institute of Nuclear Physics, 1/AF Bidhan Nagar, Kolkata 700064, India.
}

\begin{abstract}
In a recent work [{\it arXiv:1307.1382}]  it was pointed out that   the link-weight   distribution   of 
microRNA (miRNA) co-target     network  of   a wide class of  species are universal up to
scaling. The number cell types, 
widely accepted as a measure of complexity, 
turns out to be   proportional to these scale-factor.
In this article we discuss additional    universal  features of these networks and show that, 
this universality splits  if one considers distribution of number of common targets of three 
or more number of miRNAs. These   distributions for different species  can be collapsed onto 
two   distinct    set of universal functions,  revealing the  fact that the species   which appeared 
in early  evolution have   different complexity measure   compared to those appeared late.  
\end{abstract}


\maketitle

MicroRNAs are small non-coding 
single stranded RNAs   of about $22$ nucleotides long and act 
as a secondary regulator of gene  expression \cite{Liu,bartel2,ambros}.  They are  transcribed  
from the  DNA  from  either inter or intra genomic region  \cite{biogenesis,bookmir} and   bind  to the 
UTRs of   some of the  mRNAs     to inhibit  their functionality \cite{farh}.  In effect, 
the   respective proteins are produced less  compared to the  situation when 
miRNAs are absent \cite{bookmir,mir_rev}. Even  being  secondary regulators 
(transcription factors being the primary ones) miRNAs  are    seen as  potential  
therapeutic targets for treatment of cancer \cite{m_cancer,m_cancer1} and other disease \cite{m_depression,m_heart}. 
There are large number of databases \cite{miRBase,miranda} which  predict   short genomic 
sequences which might be acting as a miRNAs. 
Substantial  effort \cite{barteltarget,lewis,mirDB,TargetScan,miRWalk,microcosm} has also  been 
given in predicting the targets of these miRNAs.

The miRBase database \cite{miRBase}  predicts  that  there  are  about  $851$  miRNAs
for \textit{Homo sapiens}. A web resource 
MicroCosm Targets \cite{microcosm}  provides computationally predicted targets  
of  microRNAs across many species {\clrr (for example human miRNAs  target about  $950$  
mRNAs out of total $35864$).}  Experimental validation {\clrr of}  these predictions are, however, 
largely lacking. It is believed \cite{cooperative} that specific  biological  functions  and processes are 
possibly carried out   by  groups of miRNAs, through  noise reduction, than 
individual ones.  It is  thus important to look for combinatorial regulation \cite{combinatorial}. 
Recent   studies on miRNA \cite{mookherjee,xu} co-target network  for \textit{Homo sapiens}  reveal   the  miRNA
groups and  obtained  the  respective   functions.    The miRNA co-target network  
is formed  by   joining a  pair of miRNA by link and associating the number of co-targets 
as the    weight  of the link.   Community structures  of these  densely weighted 
networks are then obtained using certain modularization algorithms \cite{mookherjee,mirmod}.

Surprisingly the  weight distribution of these   networks  show amazing universal features, 
which extends over  {\clrr many} species classes, families and genera \cite{univ}.  In other words the  
distribution   function $P(w)$  of  the number of co-targets $w$  is  found to be a scaled  
form  of an universal  function; the species   are characterized by  a unique scale 
factor $\lambda.$   It was also observed that the scale factor is    proportional to the 
number of   cell types of  the {\clrr respective} species,  and thus,  can be considered as  a measure 
of the  complexity. A simple  random-target model, where  miRNAs  of a species target  a fixed 
number of   mRNAs, could  produce the universal function reasonably well.   Thus,     animal 
specificity is   not  resolved   through pair-wise co-targets  and    from these networks one 
expects  to obtain    generic    functions common to  species of wide range.

In this article, we  propose that,   if one  consider   number of common targets of three or more miRNAs 
of a given   species,  the  distribution function  show     {\it two } distinct groups of 
animals.  We also  observe that this  sub-groups are consistent with the  natural partition of the  
species    into   two groups obtained  from  the  bi-model distribution of  number of miRNAs.

\begin{figure}[t]
 \centering
 \includegraphics[width=8.5cm]{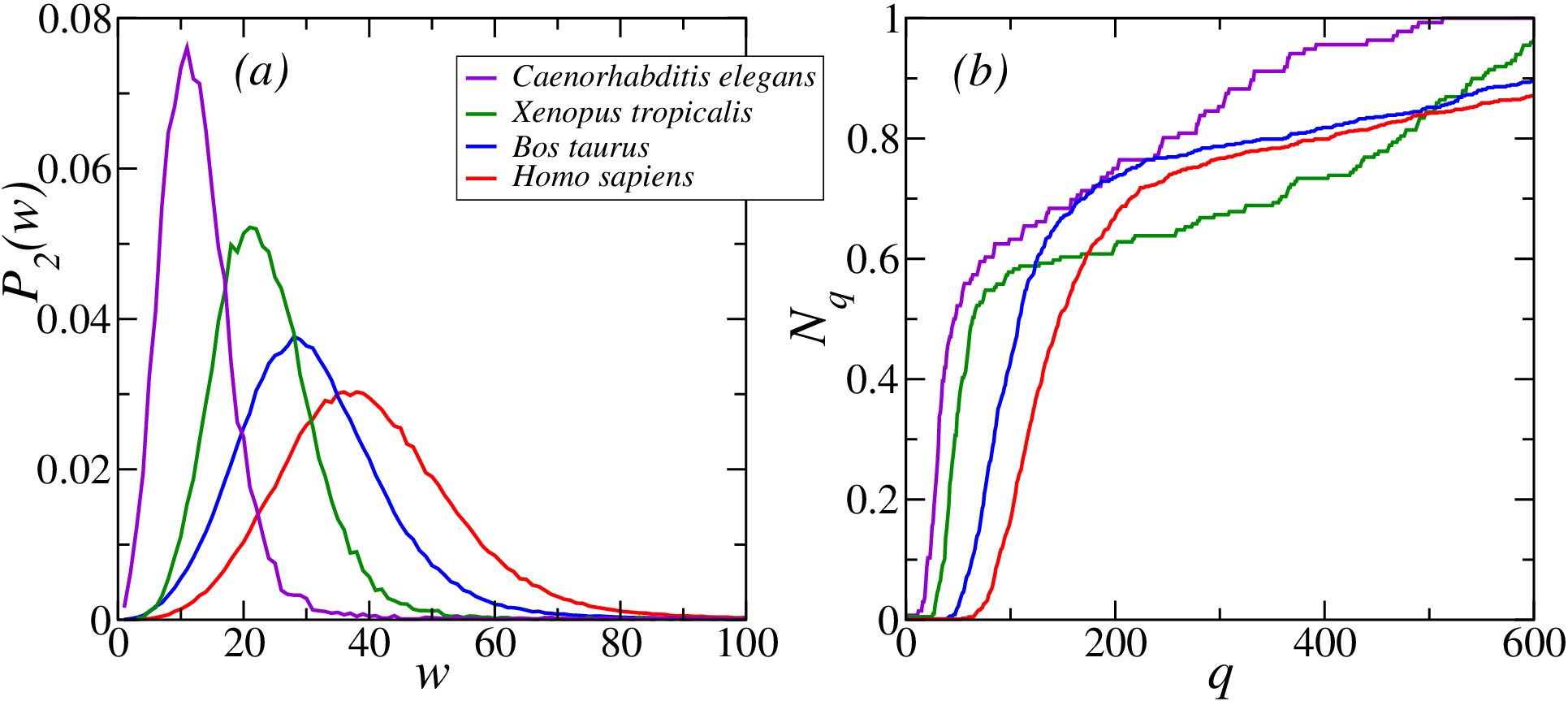}
 \caption{(a) Link-weight distribution  $P_2(w)$ and (b)  number of components $N_q$ are    
 shown for four different  species ( \textit{C. elegans}, \textit{Xenopus tropicalis}, \textit{Bos taurus} and \textit{Homo sapiens}).}
 \label{fig:Udist}
\end{figure}

 For completeness, first we discuss  how to   obtain the  miRNA co-target  distributions for 
 different species.    We consider all   the species whose miRNA  targets  have been predicted by 
MicroCosm Targets \cite{microcosm}.  The complete list of {\clrr species}, along with the number of 
miRNAs  $M$ and  mRNAs $N$ are  given in Table \ref{table:I}. Let us denote the miRNAs  
of a species  as $\{ m_1, m_2  \dots m_M\}.$    To construct   the co-target of  miRNA 
multiplets  of size $k,$  we   first   take  $k$   distinct miRNAs  $\{m_{i_1} , m_{i_2}, \dots m_{i_k}\}$
from the set of $M.$ Thus in total,  there are $C^M_k$   multiplets.  Then from the target database,   
we  find  the    targets   which are common   to  the first two  miRNAs 
 $m_{i_1}$ and   $m_{i_2}$  and denote the  number of common targets as $w_2$. These $w_2$   targets 
 are then  compared with  the targets  of   next miRNA  $m_{i_3}$  to find  $w_3$   which  is now 
 the number of common targets of   the triplet $\{m_{i_1} , m_{i_2}, m_{i_3}\}.$   
 This process is iterated until one obtains $w_k.$  The nonzero  $w_k$s  obtained    
 through  this  process are  now     considered for obtaining the  
 distribution function $P_k(w) \equiv P(w_k).$ 
 Clearly, for $k=2$,   the distribution  $P_2(w)$  is equivalent to  $P(w),$  the    
 link-weight distribution   of miRNA  co-target networks discussed in  Ref. \cite{univ}.  In  
 Fig. \ref{fig:Udist}(a)    we have  shown  $P_2(w)$ for  four different species, namely 
\textit{Homo sapiens} (Human),   \textit{Bos taurus} (bovine), 
\textit{Xenopus tropicals} (western clawed frog) 
and \textit{C. elegans}. Note that, the number of common genes   targeted  by any miRNA pair 
does not depend   on the total number  miRNAs of the species, however depends on 
the number of genes $N$.  To  understand  the    shift  in  peak position and  the change  in width  
(or variance)   of $P_2(w)$,  a simple model has been proposed in Ref. \cite{univ}. 
It was   shown   that the peak position  is  proportional to  square of  the average number of genes targeted by  the miRNAs and  inversely proportional to  $N.$  Interestingly, $P_2(w)$s  
for  different species could be   collapsed  onto each other  by aligning their peaks  at origin and rescaling   the $x$- and $y$- axis   
suitably by a  single factor $\lambda$ which  is species dependent. It was observed  that  
the scale factors  are  proportional to the  number of  cell-types and {\clrr serve as an}
 equivalent measure of  complexity of the species.

In the following we consider  certain  other  universal  features of the miRNA co-target network. 
Mookherjee \textit{et. al.} \cite{mookherjee} have  constructed the  miRNA co-target network of 
\textit{Homo sapiens}    from a  adjacency matrix $W$ with elements $w_{ij}$ 
same as the  number of   co-targets  of miRNA pair  $i$ and $j.$ 
The human   miRNA co-target network  was found to be fully 
connected with large   variation  in  link-weights - some as large as $1282$ and  as small as $1.$  
They argued that     links with small weights are   rather  un-important  and 
the network  can be made  simple  by erasing  links whose weights {\clrr are} smaller  
than   a pre-specified value $q.$ In this case, the network  breaks  {\clrr into} $N_q$  number of 
disconnected components;   $N_q$ being  a non-decreasing 
function of $q$ with $N_0=1$ (as all miRNAs are connected at $q=0$). {\clrr The variation of $N_q$ 
with $q$ for \textit{Homo sapiens} and other three species are shown in Fig. \ref{fig:Udist}(b).}   
Let us define density of components                     
\begin{equation}
 \nu_q= \frac{N_q-1}{M-1}.
\end{equation}
Evidently,   as  $q$  is increased,   {\clrr $\nu_q$} picks up a non-zero value
 at some $q=q_c$  (when  the network   starts breaking up).
Mookherjee \textit{et. al.} \cite{mookherjee} have claimed that   the optimum   network, that   does not contain 
irrelevant links nor looses   the  network    functionality, occurs 
at a   value  of  $q^*=103$  where the breaking 
 rate$\frac{d\nu_q}{dq}$    is maximum. 
The  largest component  at $q=q^*,$ which contains $429$ miRNAs,  
provide all essential regulations.  This group of miRNAs  consists of  several small 
clusters which are found to be tissue, pathway, diseases specific.  
\begin{figure}[b]
\centering
\vspace*{-.3cm}
\includegraphics[width=8.7cm]{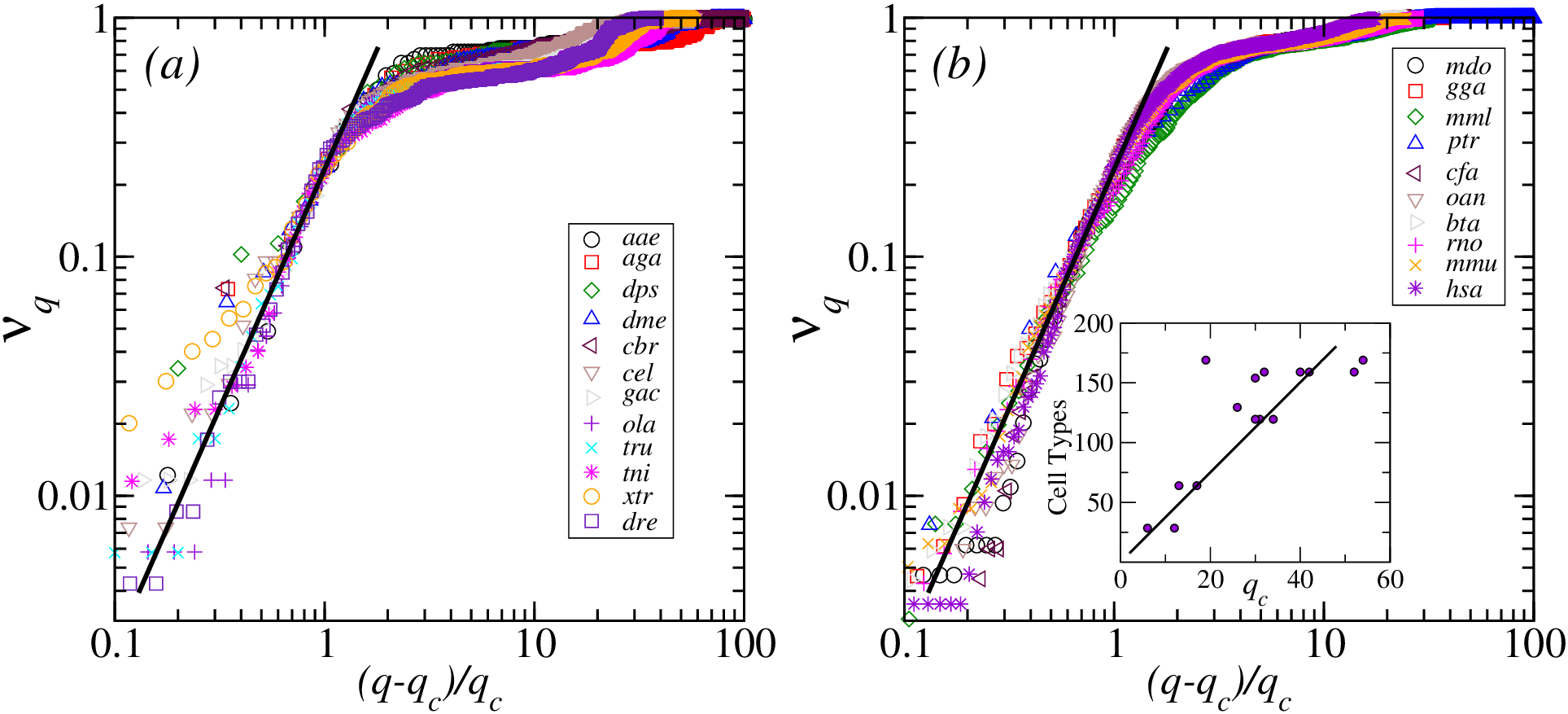}
\caption{ {\clrr $\nu_q$ vs $(q-q_c)$ for (a) Group-I and (b) Group-II, when scaled suitably,
show data collapse. The scaling function  is scale free near $q=q_c$ with exponent $\beta = 2$ (solid line). Inset of (b) shows that the $q_c$ is proportional to 
the number of cell types of the respective species; the proportionality constant, from the best fitted line,  
is $3.76.$} }
 \label{fig:univq_1}
\end{figure}

\begin{center}
\begin{table*}
  \caption{List of species and corresponding parameters.}
\label{table:I}
\hspace*{-.4cm}
\begin{tabular}{|l|c|c|c|c||l|c|c|c|c|}
\hline
{\bf Species} (short name)  Group-I & {\bf M} & {\bf N}& {\bf MYA} & $q_c$ &
{\bf Species} (short name) Group-II & {\bf M} & {\bf N}& {\bf MYA} & $q_c$\\
 \hline
 \textit{Aedes aegypti} ($aae$)  & 82 & 16059 & 285 &14 & Monodelphis domestica($mdo$) & 644 & 26013 &-&26\\
 \textit{Anopheles gambiae} ($aga$)& 82 & 12708 &   -&13& \textit{Gallus gallus} ($gga$)  & 651 & 20842&55&30\\
\textit{Drosophila pseudoobscura} ($dps$) & 88 & 12416 &-&14& Macaca mulatta ($mml$) & 656 & 32302&6.5&57\\
\textit{Drosophila melanogaster} ($dme$) & 93 & 15416 & 375&17& \textit{Pan troglodytes} ($ptr$) & 662 & 29355&2.7&19\\
\textit{Caenorhabditis briggsae} ($cbr$) & 135 & 13785 &  -&6&  \textit{Canis familiaris} ($cfa$) & 668 & 23628&5&32\\
\textit{Caenorhabditis elegans} ($cel$) & 136 & 24728 &   415&12& \textit{Ornithorhynchus anatinus} ($ana$) & 668 & 23097&115&21\\
\textit{Gasterosteus aculeatus} ($gac$) & 172 & 26423 &   -&31& \textit{Bos taurus} ($bta$) & 676 & 25759&15&40\\
\textit{Oryzias latipes} ($ola$) & 172 & 23514 &   -&25& \textit{Rattus norvegicus} ($rno$) & 698 & 30421&55&42\\
\textit{Takifugu rubripes} ($tru$) & 173 & 21972 &   -&31& \textit{Mus musculus} ($mmu$)  & 793 & 30484&55&52\\
\textit{Tetraodon nigroviridis} ($tni$) & 174 &28005 &  420&34&  \textit{Homo sapiens} ($hsa$)  & 851 & 35864&0.2&54\\
\textit{Xenopus tropicalis}  ($xtr$) &199 & 24272 &   360&26& & & &&\\
\textit{Danio rerio} ($dre$)  & 233 & 28744 &  420&30&  & &&& \\
\hline
\end{tabular}
\\ ${\bf M}:$  No. of miRNA , ${\bf N}:$  No. of  target mRNAs, {\bf MYA}: Million years ago (appeared), $q_c :$ Critical threshold.
\end{table*} 
\end{center}

We revisit   co-target  networks   for human and $21$  other species  and
find that   the  density  of  components $N_q$ also shows  certain other 
universal  features. Firstly,  all the networks are found to be fully connected, 
with    unit clustering coefficient  and diameter.
Further,  $\nu_q$    for   all    the species   show    data collapse, \textit{i.e.}  one  can  write 
$\nu_q  =  {\cal F}(A (q-q_c))$ where   $q_c$  is the critical threshold    where the network starts 
breaking   into   dis-joint components  and $A$ is a  scale factor.   We find that  $\nu_q$   is scale free 
near $q=q_c,$
\begin{equation}
 \nu_q \sim (q-q_c)^\beta. 
\end{equation}
 In Fig. \ref{fig:univq_1}    we have  plotted      
$\nu_q$ as a  function of $(q-q_c)$ in log scale  where 
(a)   corresponds to species  with   less than $250$ miRNAs and  (b) corresponds to the rest. 
The  $x$-axis is rescaled here   to obtained the collapse.  
In fact  figures (a) and (b)    could be collapsed onto each other, but 
they are  shown as  separate figures  only emphasize that  the data for   
species with small number of miRNAs   are comparatively noisy.   Our best  estimate is  
$\beta=2$; a straight line with slope $\beta=2$  is drawn   in both figures 
for comparison.  Interestingly, we find  that  critical threshold $q_c$  is also proportional to the 
number of cell-types  (see  inset of Fig \ref{fig:univq_1}(b)) and  thus, it can also be 
considered as an equivalent 
measure  of complexity.

Upto this    point,   we have discussed  that the     network formed by the common targets 
of miRNA-pairs  are universal  in many  ways: the link weight distribution $P(w)$ and 
the  density of components $\nu_q$   near the breakdown point $q=q_c$   for different species   across a
wide class  are only  scaled forms     of   respective  universal   functions.  It is rather 
surprising that   the species  specificity    show up as a scale factors.  One thus expects that  
biological   functions   co-regulated by  miRNA pairs are   possibly   less specific and  occurs 
widely across   many species.   To reveal more specific functions, which   might be   strongly 
species   dependent, we  try to   find out    common targets of   more number of miRNAs, by taking $k>2.$
Note that   for $k>2$  the   number of common targets  can  not be simply interpreted as the the   weights 
of some network  (as they carry    three or more index  referring generically to tensors  which, unlike  matrices, 
does not have  a network representation).  In the following we  study  the distribution  of number of 
common {\clrr targets}  for  $k>2$ number of miRNAs. 

 \begin{figure}
 \centering
 \includegraphics[width=8.7cm]{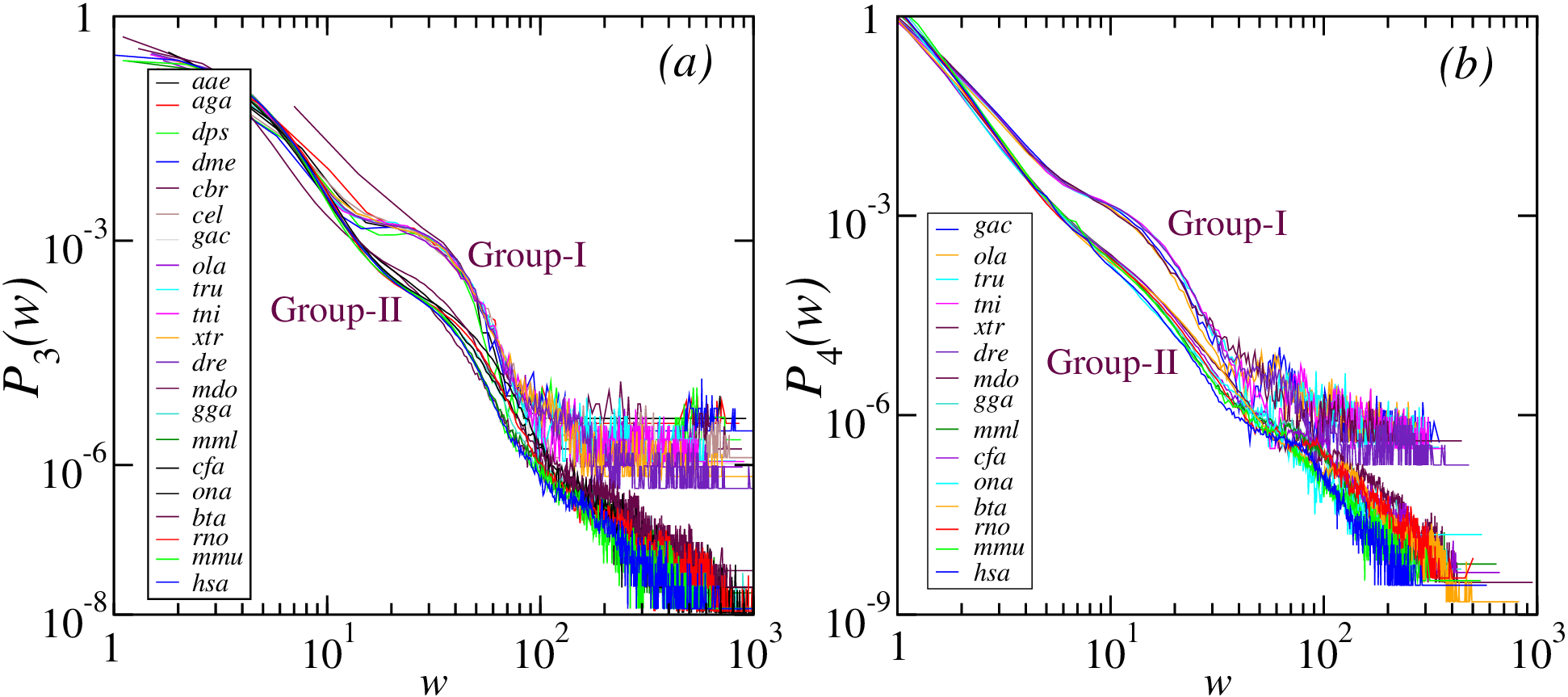}
 \caption{  (a) Distribution  functions  (a) $P_3(w)$   and (b)  $P_4(w)$    of  number
 of miRNA co-targets   for different species. In (b),    first $6$   species  of group-I  are 
 not  shown as the data is noisy  (for  $M$ being  small). The $y$ -axis   in both  plots  
 are  scaled  here for obtaining the data-collapse.}
 \label{fig:univq}
\end{figure}

 First   $k=3.$    For each species  there are   $C^M_3$ triples  and the number   of common  targets of 
  any three   different  miRNAs $i$, $j>i$ and $k>j$    is denoted as $w_{i,j,k}.$  We  find out these 
  numbers    from the   miRNA target database \cite{microcosm},  using a in house   code   and   obtain  the 
  distribution  of these numbers, denoted as $P_3(w).$    The same way one can obtain the distribution 
  $P_4(w)$ of     number of common targets of  $k=4$  miRNAs.     These distribution functions are 
  shown in Fig. \ref{fig:univq}(a) and (b)  respectively.  In these {\clrr plots} only the $y$ axis is 
  scaled   for    obtaining a data collapse. However all the data could not be collapsed on a  
  single function;  rather they split  into two  different scaling functions.   The splitting is 
  clearly visible   for $P_3(w).$  For $P_4(w)$   we find that  the data  is  too noisy for 
  species with small number of   miRNAs (first $6$ species  of group-I, in  table \ref{table:I})  
  and it was not clear whether  they   just represent noise   due to small number of miRNAs and targets, 
  or there are further sub-classes (splitting).   
  We  have not   shown  $P_4(w)$ for  these   species   as they obstruct visibility of the  other two     collapsed-curves. {\clrr Note that  the plot also does not contain the distribution functions for
  \textit{Pan troglodytes} as its number of targets is 
much lower compare to other species having nearly same number of miRNAs. }

  Why do we see universality splitting   for    mRNAs  which are targeted by larger number of miRNAs ?
  Those mRNAs  which   can be regulated by more number of miRNAs 
   take part  in  larger   number  of    biological functions or pathways   providing  possibility 
   of more complex gene-regulation.    In this regard, it  is  quite possible that  universality splitting 
   reflects  these  complexity.  The complexity  of  species belonging to  one specific  scaling 
   function  could be   strikingly different   from those belonging to the other scaling function.
   To verify, if this is indeed the case, in the following, we try to find out other measures which show
   the same kind of division.

\begin{figure}[h]
 \centering
 \includegraphics[width=7.8cm]{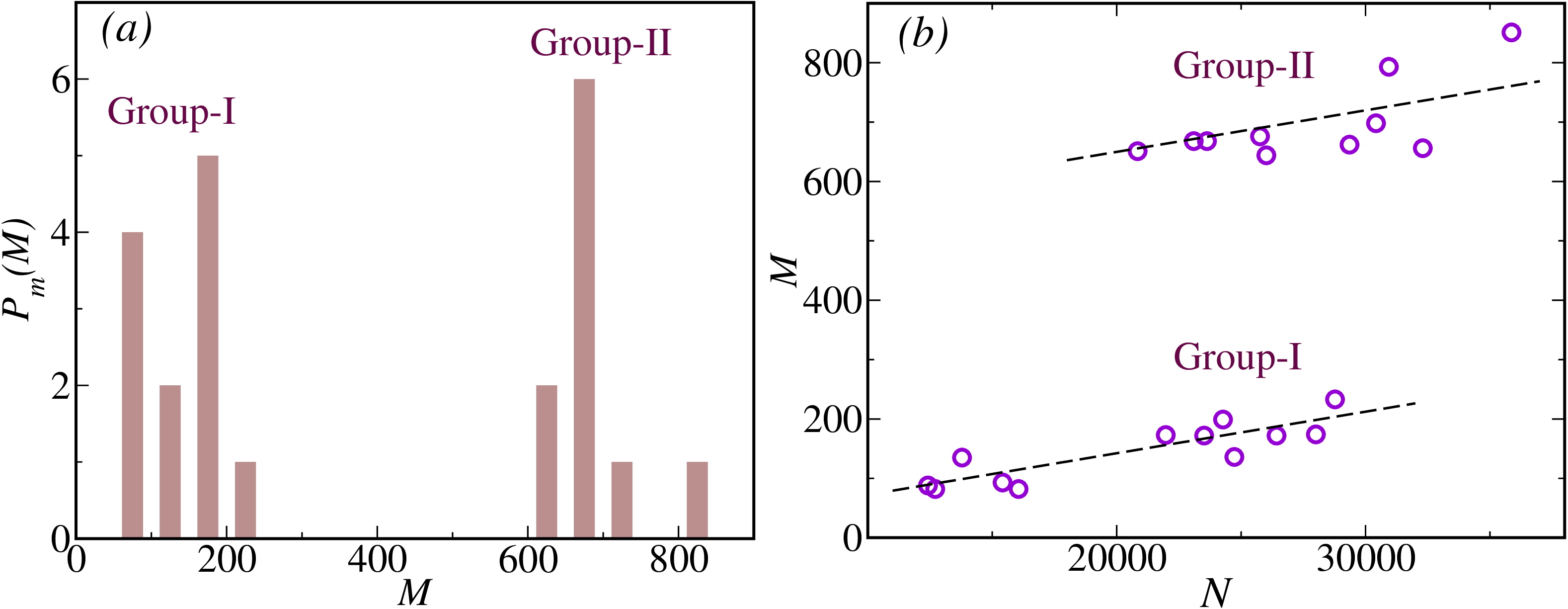}
 \caption{{\clrr (a) Distribution of number of miRNAs $P_m(M).$ (b) The numbers of miRNAs $(M)$ as a function of the number of targets $(N).$  Two distinct peaks in (a) and two different linear relations
 in (b),  with same slope $0.007$ (within error limits) but different intercepts $2.3$ and $509.8$ obtained from the  best fitted line,   are clear indications that   the  species under investigation    form   two different groups  with  respect to the number of miRNAs  they have.}} 
 \label{fig:histo}
\end{figure}

First we look at the distribution of number of miRNAs  $P_m(M).$    {\clrr Since, there are only 
a few species,  we represent the distribution by a histogram by  counting the number  of species  
 having  $M-25$  to $M+25$  species for   $M=25, 75,\dots,825$) (refer to Fig. \ref{fig:histo}(a))}.     Clearly they show 
 {\it two}  distinct  peaks  centered about $175$ and  $700.$ 
It is   rather natural  that less complex species, those who  originate in early evolution (see Table \ref{table:I}),  
have smaller  number of   miRNAs  compared to those which are more complex. But the reason  
for  the  bi-modal structure in   $P_m(N)$  is   not clear.  
Again, existence of these two groups can also be seen   from  the plot of total number 
of miRNAs $M$   versus $N$  in Fig. \ref{fig:histo}(b).  Here  $M$ varies linearly with total number 
of {\clrr target mRNAs} $N$ with a slope  $0.007$; however the $y$-intercept  for group-I ($2.3$)  
is different from that of group-II ($509.8$).  The natural division of these groups   are 
consistent with the group of species   belonging  to   two  different scaling functions.  
In other words,   one may say that the   universality splitting   is an indication that  there are 
intrigue regulatory mechanisms   associated with  
more complex species.

\begin{figure}[h]
 \centering
\vspace*{.2cm}
\includegraphics[width=8.5cm]{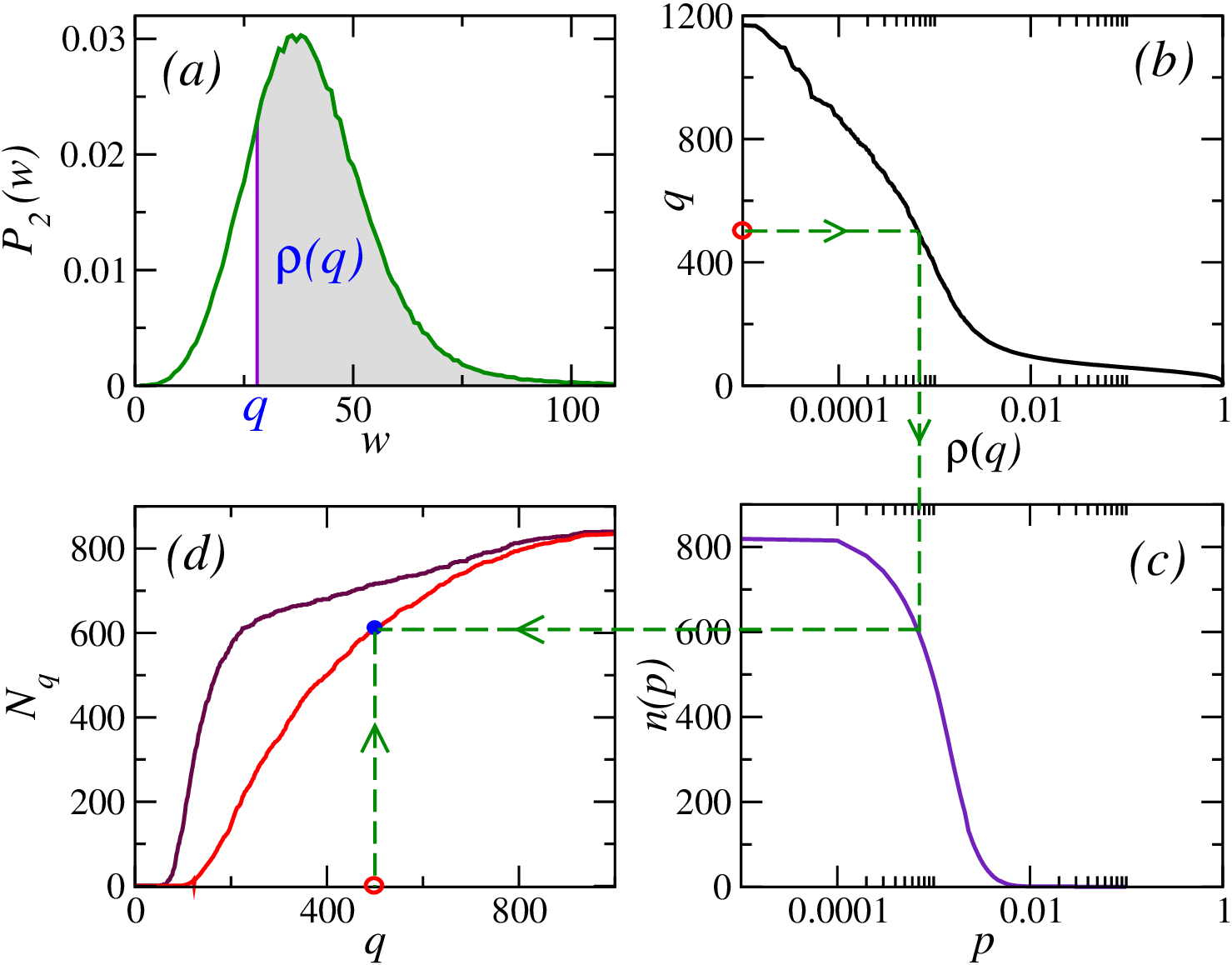}
 \caption{(a)  Link-weight distribution  $P_2(w)$  for \textit{Homo sapiens} ($M=851$) 
 integrated  for  $w>q$  
 and the resulting  $\rho(q)$ is  plotted  in (b). (c) Number of  disjoint components $n(p)$ 
 for  a  random   network  ($851$  nodes and connection probability  $p).$  (d) 
 $n(\rho(q))$  for the random network is compared with $N_q.$  Dashed-lines here 
 describes how to obtain $n(\rho(q)) $ for  a given $q$ (see text for details).  }
 \label{fig:compare_Nq}
\end{figure}

It is thus natural to ask     whether these complexity structures  are  also  hidden somewhere 
in  the    co-target network of miRNA {\it pairs}.   In the following we show that,  
even though   $\nu_q$  show   universal features, it does not capture all the 
underlying  correlations  of the network.   This can be done  by comparing   the    number of 
components  of a species with an equivalent random graph.   First note that $N_q$  and $P_2(w)$ are 
related  and  $N_q$ can  be calculated     from $P_2(w).$  Let     
\begin{equation}
 \rho(q)=\int_{q}^{\infty} P_2(w)dw
\end{equation}
be  the  cumulative distribution  of  $P_2(w),$  which is same as the   probability that 
 the  link-weight is larger than $q.$  Since the   number of components  $N_q$ is obtained 
by  erasing all the links  with    weight $w_{ij}<q$ and  assigning   unit weight to all other 
links ($w_{ij}>q$),  the  effective network   has link density  $\rho(q).$   We construct  a 
random network   of $N$ nodes   where the  pairs  are connected with   probability  
$\rho(q)$    and compare the number   of  disjoint components    of this network    with 
$N_q.$  Let the average number of   components of a random network    with connection probability  $p$  and 
number of nodes  $N$ nodes  be   $n(p).$    Then,  if  the   miRNA co-target networks were  
uncorrelated,    it is expected that  
\begin{equation}
 N_q = n(\rho(q)).
\end{equation}
In   Fig. \ref{fig:compare_Nq}(d) we  plot $N_q$ for   \textit{Homo sapiens} ($N=851$)  
along with  $n(\rho(q)).$   The construction  procedure is demonstrated    
in this  figure.   The $P_2(w)$ for \textit{Homo sapiens} ( Fig.  \ref{fig:compare_Nq}(a)) is integrated 
for $w>q$  (shaded region) to obtain $\rho(q)$    ( Fig.  \ref{fig:compare_Nq}(b)).   
Figure  \ref{fig:compare_Nq}(c)   shows   $n(p)$  for a random network   of $N=851$ nodes. 
To obtain $N_q,$  one starts  from a  given $q$ (shown as an open circle)  in 
Fig.  \ref{fig:compare_Nq}(b), obtain  $\rho(q)$  and  read out  the   number  of   components   from 
(c)   following   the  dashed line,  and  get   the data point $(q, n(\rho(q))$  which is shown as
a solid circle in Fig. \ref{fig:compare_Nq}(d).  Repeating this   for different  values  of $q$ 
we obtain the expected  $N_q$ (red-line  in Fig. \ref{fig:compare_Nq}(d))   for a random network.   Clearly,  this curve  is 
substantially different  form  the actual  $N_q$  versus  $q$   curve  for \textit{Homo sapiens},  
indicating presence of correlation in miRNA co-target network. 
Thus the  co-target network is not  just another random network   with  a specific  link-weight
distribution  $P_2(w).$  These hidden correlations   are   uncovered   in a way  when one  
considers   co-targets of three  or  more miRNAs. 

In conclusion we have studied     the distribution of number of   co-targets of  $k$ number 
of miRNAs. For $k=2$  these  numbers  can be interpreted as the    link-weight distribution 
$P_2(w)$ of miRNA co-target network, which is  known to have universal features. In this 
case, $P_2(w)$s   for different species   are only a scaled form  of an universal scaling 
function, and the scale-factor is a measure of complexity.   {\clrr We show that when links of small 
weights  (less than $q$)  are erased these networks  breaks  into several components. 
At the  breakdown point we find an additional universal feature; the number of components 
show a scale free behaviour $N_q\sim (q-q_c) ^2$ and could be collapsed onto each other 
by rescaling of only $x$-axis.}   For $k>2$ , the 
number of co-targets does not have an  graphical representation, but their distribution could 
also be  collapsed.   Surprisingly, $P_k(w)$ for $k>2$ studied for  $22$  species show universality splitting, \textit{i.e.}  $P_k(w)$  for a  group  of species collapse  onto  one scaling function 
where as  the others   belong to  a different scaling function. The universality spiting is consistent  
with the   bi-modal distribution of  number of miRNAs and the linear dependence of  number of 
miRNAs on the number of mRNAs ( two groups have different $y$-intercept). The  two different scaling
 functions are thus associated with two different  class of animal, the early ones 
which  have less number of miRNAs  and  less complex and the  late ones which are more complex. 
It remains  to study, if complex regulation occurring due to the fact that if mRNAs, which are 
targeted by more number of miRNAs  contribute to the  possibility of complex regulation and new 
biological functions.

\end{document}